\documentclass[final,5p,times,twocolumn]{elsarticle}
\usepackage[T1]{fontenc}
\usepackage{amsmath,amssymb,amsfonts,booktabs,graphicx}
\usepackage[hidelinks]{hyperref}
\journal{Physics Letters B}
\newcommand{\Mpl}{M_{Pl}}

\newcommand{\dd}{\mathrm d}
\newcommand{\GeV}{\mathrm{GeV}}
\newcommand{\phik}{\widetilde\phi}
\newcommand{\Cbar}{\overline C_L}
\newcommand{\HH}{\mathcal H}

\newcommand{\avg}[1]{\left\langle#1\right\rangle}
\graphicspath{  {images/} }

\begin{document}
\begin{frontmatter}
\title{UV/IR mixing from higher-order interactions in a scalar field}
\author{Satish Ramakrishna}
\affiliation{organization={Department of Physics \& Astronomy, Rutgers, The State University of New Jersey},addressline={136 Frelinghuysen Road},city={Piscataway},postcode={08854-8019},state={NJ},country={USA}}
\address{\normalfont\href{mailto:ramakrishna@physics.rutgers.edu}{\texttt{ramakrishna@physics.rutgers.edu}}}
\begin{abstract}
The observed vacuum energy lies far below the usual quantum-field-theoretic estimate. We construct a preferred-frame reduced scalar Hamiltonian in which a nonlocal quartic field derivative interaction produces a wavevector dependence proportional to $|\vec k|^8$ for modes below a smoothly defined scale. Within the quartic oscillator regime, the ground-state energy and excitation gaps are $\propto |\vec k|^{8/3}$. The preferred frame is characterized by a timelike clock-field background.  The Planck energy is reached at a substantially smaller value of the spatial wavevector due to its supra-linear mode-energy dependence. The proposed interaction is instantaneous in the preferred frame and yields a Hamiltonian bounded below. Normalizing the peak of the ground-state energy to the Planck scale, together with Cohen-Kaplan-Nelson (CKN) saturation, applied as a matching recipe rather than as an {\it {ab initio}} calculation, fixes the cutoff scale at $k_G\sim k_{Pl}^{1/3}k_{box}^{2/3}$. Modes above the smooth cutoff scale retain their ordinary scalar dynamics. We assume that the interaction-induced ground-state energy shift gravitates, as do other contributions. CKN saturation is imposed on the total vacuum energy, comprising this shift and the other vacuum contributions. As an alternative vacuum energy prescription, we considered using a depletion of the density of states {\it \`a la} Banks \& Draper. Used as a count of single-particle states, that approach leads to fewer electron states than there are electrons in the universe - it is hence not used, in favor subtracting the free zero-point baseline vacuum-energy by normal ordering.  Nonlocal interactions induce the possibility of superluminal information transfer. We compute the spacelike commutator of the scalar field and show that it is very small, about $10^{-69}$ of the free commutator inside the light cone. We outline how this scalar field could couple to gauge fields through a gauge-invariant radial Higgs extension. Coupling to gravity is also discussed. Identifying the inferred infrared scale phenomenologically with the cosmic event horizon reproduces the holographic-dark-energy relation $\rho_{vac}\propto\frac{M_{Pl}^2}{R_h^2}$, provided the coupling tracks the horizon; the resulting holographic fluid has $w_0\simeq-0.89$.
\end{abstract}
\end{frontmatter}

\section{Introduction}
\label{sec:intro}

The zero-point sum over wavenumbers for relativistic fields that is cut off at the Planck scale gives a vacuum-energy density many orders of magnitude above the observed dark-energy density~\cite{Jerome}. A more careful calculation that preserves Lorentz invariance and the appropriate equation of state reduces the size of the computed value, but it is still many orders of magnitude above the observed value. Weinberg's theorem further constrains local QFT mechanisms in a translationally invariant vacuum~\cite{Weinberg} to disallow a small cosmological constant without fine-tuning. One possible way around this impasse is to abandon locality; nonlocal approaches to the cosmological-constant problem have been studied in several settings~\cite{Sean}, while UV/IR mixing is also familiar from noncommutative field theory and gravitational consistency bounds~\cite{Minwalla,CKN1,CKN2}.

An alternative approach is taken with the Cohen-Kaplan-Nelson bound. That bound requires the energy in a region of linear size $L$ not to exceed the mass of a black hole of the same size,
\begin{equation}
L^3\rho_{vac}\lesssim LM_{Pl}^2
\qquad
\rho_{vac}\lesssim \frac{M_{Pl}^2}{L^2}
\label{eq:CKNintro}
\end{equation}

The bound is a consistency condition on the total energy in a region, including the energy of any ordinary or dark matter present~\cite{CKN1,CKN2,CKN3,CKN4}. On its own, it does not specify a field-theoretic mode spectrum or explain why the dominant contribution to the vacuum-energy shift should come from modes far below $k_{Pl}$. In our vacuum-energy calculation we impose saturation for a region in its ground state, with no matter excitations considered.

In this paper, we construct a reduced scalar Hamiltonian with an additional quartic coefficient that additionally contains an explicit parametrized Gaussian envelope
\begin{equation}
\lambda_k=\zeta L^3k^8G(k)\qquad G(k)=e^{-(k/k_G)^2}
\label{eq:lambdaintro}
\end{equation}
The spatial scale $k_G$ is an interaction scale and will naturally produce a smooth cutoff. The $k^8$ factor gives a $k^{8/3}$ ground-state energy spectrum only where the quartic dominates and the envelope changes slowly. The full quadratic-plus-quartic oscillator energy spectrum returns to ordinary massive behavior as $G(k)$ suppresses the quartic interaction - a roton-like dispersion. A useful consequence of this is that we are able to revert to a full spectrum of allowed wavevectors above the $k_G$ interaction scale. The field theory is an effective one, valid below the Planck wavevector. We deal with the contribution to the vacuum energy from the linearly dispersive part of the spectrum by requiring a normally ordered Hamiltonian. By definition, therefore, the vacuum energy only gets contributions from loop corrections, potential minima and these non-linear terms.

We fix parameters as follows: we set the Planck energy for the computed ground-state energy peak and posit CKN saturation for the vacuum energy (though an alternative matching condition is discussed in Section 7). Assuming a fixed Gaussian envelope and the CKN saturation being applied at every epoch during the universe's evolution,
\begin{equation}
k_G\sim k_{Pl}^{1/3}k_{box}^{2/3} \: \: \: \: \: \: \: \: \text{where} \: \:
\qquad k_{box}=\frac{2\pi}{L}
\qquad k_{Pl}=M_{Pl}
\label{eq:mainclosureintro}
\end{equation}
The construction relies on two separate ideas.
\begin{enumerate}
\item The first is this - the parametrized Gaussian multiplies the interaction in a quartic Hamiltonian. In position space it averages each field over a distance of order $\frac{1}{k_G}$ (see Sec.~\ref{sec:ssb}). It removes no modes from the Hilbert space and leaves the canonical commutator unchanged.
\item The second idea provides a method for states of wavevectors higher than the quartic's cutoff to exist. We assume that the full interaction applies and the quartic term dominates for small wavevectors, while the zero-point energy of the free modes is removed by normal ordering (Sec.~\ref{sec:prescription}).
\end{enumerate}
No particle states are removed. Subtracting the free zero-point baseline does not change the free field's excitation energy spectrum. The quartic does change them inside the band, and the ordinary energies are obtained outside it. The CKN bound is saturated by the vacuum energy. The Gaussian is nonzero at every finite momentum; hence the excess over the quadratic terms must be integrated over all momenta and shown to be finite and fixed by matching with the CKN bound.

The nonlocal interaction opens the possibility of superluminal signalling; we compute the spacelike commutator and find it suppressed by $\big(\frac{k_e}{m}\big)^3$, where $k_e$ is the upper edge of the quartic band. The matching and this computation are carried out below (Secs.~\ref{sec:excess} and~\ref{sec:causality}).

The broader UV/IR literature includes holography~\cite{tHooft,Susskind,SusskindWitten}, noncommutative field theory~\cite{SeibergWitten,Minwalla,Matusis,Abel2006}, and other constructions connecting separated scales~\cite{ArkaniHarigaya,Kephart,Abel2026}. Preferred-foliation and clock-field constructions provide the setting for the reduced dynamics~\cite{ArkaniHamed}.

\section{Lorentz invariance and supra-linear pole laws}
\label{sec:lorentz}
In a translationally invariant vacuum with a Lorentz-invariant self-energy $\Sigma$, this self-energy depends on the external momentum only through $u=\omega^2-\vec k^2$. The dressed pole condition $\omega^2-\vec k^2-m_0^2-\Sigma(\omega,\vec k)=0$ then becomes
\begin{equation}
 u=m_0^2+\Sigma(u)\qquad \omega^2=\vec k^2+u_\star
 \label{eq:pole}
\end{equation}
with roots $u_\star$ independent of $\vec k$. Every pole branch is therefore asymptotically linear, and within the standard single-pole description a supra-linear pole law needs dependence on $\vec k^2$ separately from $\omega^2$, i.e.\ a preferred-frame sector. While this does not exclude other approaches, it motivates the route adopted below.

\section{Preferred-frame nonlocal interaction}\label{sec:ssb}
\subsection{Preferred time, Fourier conventions and interaction}

We use $\hbar=c=1$, signature $(+,-,-,-)$, and consider a real scalar of mass $m>0$. We assume that the vacuum selects a timelike unit vector $n^\mu$. A convenient realization is a derivative-coupled clock field $T$ \cite{ArkaniHamed}, which spontaneously condenses as below
\begin{equation}
n^\mu=\frac{\partial^\mu T}{\sqrt{\partial_\alpha T\partial^\alpha T}}
\qquad \langle T\rangle=M^2t
\label{eq:clock}
\end{equation}
and $n^{\mu}$ is defined only in this phase, since $\partial_\mu T$ vanishes in the symmetric one. The action is Lorentz covariant and the vacuum spontaneously breaks boosts~\cite{ArkaniHamed}. Only the existence of the preferred frame is used below. Define (we use $g_{\mu \nu}$ for the Minkowski metric tensor)
\begin{equation}
P_{\mu\nu}=g_{\mu\nu}-n_\mu n_\nu
\qquad
D_n=P^{\mu\nu}\partial_\mu\partial_\nu
\end{equation}
In the preferred frame, $n^\mu=(1,0,0,0)$ and $D_n=-\nabla^2$. A quartic operator $(D_n\phi)^4$ therefore gives two powers of spatial momentum on each leg and a four-field vertex proportional to $|\vec k|^8$.
 
In a periodic region $V=L^3$, we define some Fourier transforms of the real-space field variable as
\begin{align}
 \phik(\vec k,t)&=\int_V\dd^3x\,e^{-i\vec k\cdot\vec x}\phi(\vec x,t)\nonumber\\
 \phi(\vec x,t)&=\frac1V\sum_{\vec k}\phik(\vec k,t)e^{i\vec k\cdot\vec x} \: \: \: \: 
 \qquad \vec k=\frac{2\pi}{L}\vec j
 \label{eq:fourier}
\end{align}
where $\vec j\in\mathbb Z^3$, the set of integers.
Thus the mass dimensions are $[\phi]=\GeV$, $[\phik]=\GeV^{-2}$ and $\phik(-\vec k)=\phik(\vec k)^*$. 
 
To motivate the Gaussian envelope mentioned in Sec.~\ref{sec:intro}, we define a spatially averaged field (in flat space) only for use in the proposed extra interaction as follows,
\begin{align}
 \phi_G(\vec y, t) = \frac{k_G^3}{\pi^{3/2}}\int \dd^3 x \:  e^{-k_G^2 (\vec y-\vec x)^2} \: \phi(\vec x,t) \: \: \: \:  \: \: \: \:  \nonumber \\
 \phik_G(\vec k)=f_G(k)\, \phik(\vec k) \: \:  f_G(k)=e^{-k^2/(4k_G^2)} \: \:  f_G(k)^4=G(k)
 \label{eq:fieldfactor}
\end{align}
The canonical kinetic and mass terms continue to act on $\phi$. Only the interaction piece uses $\phi_G$. The total  Hamiltonian, which is written as the sum of the free Klein-Gordon piece and an interaction piece, is
\begin{align}
 H&=H_{KG}+U_4\nonumber\\
 H_{KG}&=\frac1{2V}\sum_{\vec k}\big[|\widetilde\Pi(\vec k,t)|^2+(k^2+m^2)|\phik(\vec k,t)|^2\big]\nonumber\\
 U_4(t)&=\frac\zeta V\sum_{\vec k}k^8e^{-(k/k_G)^2}|\phik(\vec k,t)|^4\nonumber\\
 &\qquad\qquad \zeta>0\qquad [\zeta]=\GeV^{-2}
 \label{eq:Hdiag}
\end{align}
where $H_{KG}$ is the free Klein-Gordon Hamiltonian, {\it {i.e.}}, $\frac12\int_V\dd^3x\,[\Pi^2+(\nabla\phi)^2+m^2\phi^2]$, with conjugate momentum $\Pi=\partial_t\phi$, $[\phi(\vec x,t),\Pi(\vec y,t)]=i\delta_L^{(3)}(\vec x-\vec y)$ and the Fourier transform $\widetilde\Pi$ is computed as in Eqn.~\eqref{eq:fourier}. The sums include all momenta. The Gaussian multiplies the $k^8$ quartic coefficient once; equivalently it is the fourth power of the per-field factor in Eqn.~\eqref{eq:fieldfactor}. We note (again) that modes in the high-momentum tail become progressively harmonic and are not removed from the state space. 
 
$H_{KG}$ carries the ordinary zero-point energy, which is removed by normal ordering. The vacuum energy includes the ground-energy shift produced by $U_4$ and certain other contributions, explained in Sec.~\ref{sec:prescription}.
  
Let us analyze where such an interaction term could come from. In real space the same interaction is
\begin{equation}
 U_4=\zeta\int\prod_{i=1}^4\dd^3x_i\,
 \delta_L^{(3)}(\vec x_1+\vec x_4-\vec x_2-\vec x_3)
 \prod_{i=1}^4(D_n\phi_G)(\vec x_i,t)
 \label{eq:realspace}
\end{equation}
When we consider this in Fourier transform space, orthogonality selects the opposite-momentum pairs; the four spatial derivatives produce $k^8$ and the four factors $f_G$ give $G(k)$. 
Here the Gaussian average is applied to each field $\phi$. Equivalently, the fields can be left unaveraged and
$\delta_L^{(3)}(\vec A)$ in Eqn.~\eqref{eq:realspace}, with $\vec A=\vec x_1+\vec x_4-\vec x_2-\vec x_3$, replaced by the single
function $\mathcal K_G(\vec A)=\frac{k_G^3}{8\pi^{3/2}}\,e^{-k_G^2A^2/4}$, the convolution of the four averaging Gaussians, whose
Fourier transform is $G(k)$. 
 
This is a nonlocal four-point interaction: small $A$ does not require every pair of points to be close. The Gaussian is an infinite spatial-derivative operator - it is bounded by $1$ and acts only along the preferred-time slices, so the difficulties of Lorentz-invariant form factors $e^{-\Box/M^2}$~\cite{Carone} do not arise.
 
The coefficient $\zeta$ is not derived from a microscopic theory. Its size is motivated by a spacetime-nonlocal version of the quartic, in which the four fields are joined by a dimensionless, non-singular and rapidly decaying function $C$:
\begin{align}
 S_4={}&-g\int\prod_{i=1}^4\dd^4x_i\,\delta^{(4)}(x_1+x_4-x_2-x_3)\prod_{i=1}^4(D_n\phi)(x_i)\nonumber\\
 &\times C(x_1-x_2)\,C(x_1-x_3)\,C(x_2-x_3)\,C(2x_1-x_2-x_3)\nonumber\\
 C(s)={}&e^{-A^4|s|_E^4}
 \label{eq:S4}
\end{align}
Here $g$ is dimensionless, $A$ is an inverse interaction range and $|s|_E$ is the Euclidean length of the four-vector $s$ in the preferred frame, so $0<C\leq1$. In a cell of period $L$ in time and in each spatial direction, define the function's average and the matched coefficient
\begin{align}
 \Cbar&=\frac1{L^4}\int_{[-L/2,L/2)^4}\dd^4s\,C(s)\nonumber\\
 \zeta&=gL^2\Cbar^{\,4}\qquad 0<\Cbar\leq1
 \label{eq:zetamatch}
\end{align}
A static match over a temporal period $L$ gives Eqn.~\eqref{eq:zetamatch}, with $\Cbar\simeq1$ for $AL\ll1$. It fixes only the size of the coefficient: Eqn.~\eqref{eq:Hdiag} is the definition of the model that we use. The matching condition is itself an ansatz, introduced to motivate the size of $\zeta$, as we will discuss in Sec.~\ref{sec:matching}. It does not fix $G(k)$, which is a second, independent ansatz.

\subsection{Spectrum and stiffness}
\begin{figure*}[t]
\centering
\includegraphics[trim=0 16bp 0 0,clip,width=0.88\textwidth]{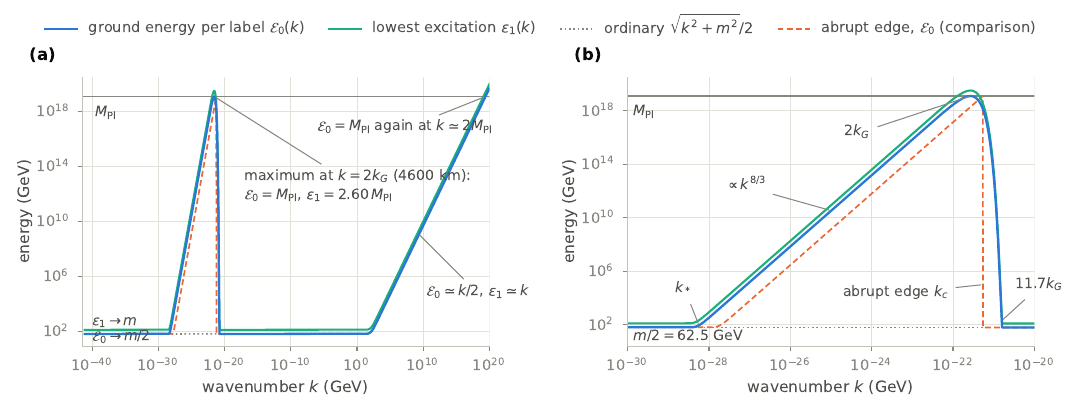}
\caption{Ground-state energy per Fourier mode $\mathcal E_0$ and lowest excitation gap $\epsilon_1$ of the pair Hamiltonian in Eqn.~\eqref{eq:pair}, at the benchmark of Sec.~\ref{sec:excess}. (a) Both rise as $k^{8/3}$ for $k_*\ll k\ll k_G$ and reach a local maximum near $2k_G$, where $\mathcal E_0\simeq\Mpl$ and $\epsilon_1\simeq2.60\Mpl$. At the upper band edge $k_e\simeq11.7k_G$, they are $1.48$ and $1.76$ times their free values, respectively. Beyond the band they return to their mass plateaus. For $k\gg m$ the free branches grow as $k/2$ and $k$. Extrapolating the ground-energy branch beyond the range of the effective theory gives $\mathcal E_0\simeq\Mpl$ again at $k\simeq2\Mpl$. (b) A close-up of the quartic band and its edges, $k_*\simeq6\times10^{-29}$~GeV and $k_e$. The dashed curve shows $\mathcal E_0$ for an abrupt band edge with the same integrated vacuum-energy shift.}
\label{fig:E0}
\end{figure*}

Each nonzero conjugate pair $(\vec k,-\vec k)$ in Eqn.~\eqref{eq:Hdiag} decouples. For one member $\vec k$ of each pair write $\phik(\vec k,t)=\sqrt{\frac V2}\,[x_k(t)+iy_k(t)]$ and $\widetilde\Pi(\vec k,t)=\sqrt{\frac V2}\,[p_{x,k}(t)+ip_{y,k}(t)]$, so that $[x_k,p_{x,k}]=[y_k,p_{y,k}]=i$. Since $\phik(-\vec k)=\phik(\vec k)^*$, the terms $\pm\vec k$ of Eqn.~\eqref{eq:fourier} add to $\frac{2}{V}\operatorname{Re}[\phik(\vec k,t)e^{i\vec k\cdot\vec x}]$, so
\begin{equation}
 \phi(\vec x,t)=\frac{\phik(0,t)}V+\sqrt{\frac2V}\sum_{pairs}\big[x_k\cos(\vec k\cdot\vec x)-y_k\sin(\vec k\cdot\vec x)\big]
 \label{eq:pairfield}
\end{equation}
In both sums of Eqn.~\eqref{eq:Hdiag} the terms $\pm\vec k$ are equal, and each quartic one is $\frac{\lambda_kS^2}{4}$, with $S=x_k^2+y_k^2$. Dropping the label $k$,
\begin{align}
 H_k&=\frac{p_x^2+p_y^2}{2}+\frac{k^2+m^2}{2}S+\frac{\lambda_k}{2}S^2\nonumber\\
 \lambda_k&=\zeta Vk^8e^{-(k/k_G)^2}
 \label{eq:pair}
\end{align}
a two-dimensional oscillator with a confining quartic, self-adjoint and bounded below. Its first two terms form the free pair Hamiltonian $H_f$. In the quartic regime, where we ignore the quadratic term, we scale $x,y=\lambda_k^{-1/6}(X,Y)$, preserve the commutators and find $H_k = \lambda_k^{1/3}h_Q$ with $h_Q=\tfrac12(P_X^2+P_Y^2)+\tfrac12(X^2+Y^2)^2$. The ground energy per Fourier component is hence a pure number times $\lambda_k^{1/3}$. By this simple scaling argument, with $\gamma_4$ determined numerically, we obtain
\begin{equation}
 \mathcal E_Q(k)=\gamma_4(\zeta V)^{1/3}k^{8/3}e^{-k^2/(3k_G^2)}
 \qquad \gamma_4\simeq0.5862
 \label{eq:E0}
\end{equation}
We diagonalize Eqn.~\eqref{eq:pair} numerically at every $k$. This gives the exact ground energy per label, $\mathcal E_0(k)$, and the lowest excitation the field can create, $\epsilon_1(k)$. Both depend on $k$ only through $\Omega_k=\sqrt{k^2+m^2}$ and $\eta_k=\frac{\lambda_k}{\Omega_k^3}$, and they reduce to $\frac{\Omega_k}{2}$ and $\Omega_k$ for $\eta_k\ll1$. Figure~\ref{fig:E0} shows the result, which is similar to a roton dispersion curve: a single maximum at $k=2k_G$, a wavelength of about $4100$~km, on the mass gap, with no minimum below $m$.

The stiffness of a mode is measured by $\eta_k=\frac{\lambda_k}{\Omega_k^3}$, the ratio of the quartic to the quadratic energy at the amplitude of the field in the quadratic ground state $\avg S=\frac{1}{\Omega_k}$. In the band $k\ll m$, so $\Omega_k\simeq m$ and $\eta_k\simeq\frac{\lambda_k}{m^3}$. Normalizing the quartic-regime peak in Eqn.~\eqref{eq:E0} by $\mathcal E_Q(2k_G)=\Mpl$ gives
\begin{equation}
 \eta_k=\Big(\frac{\Mpl}{\gamma_4m}\Big)^3\Big(\frac{k}{2k_G}\Big)^8e^{4-(k/k_G)^2}
 \label{eq:eta}
\end{equation}
The prefactor is the value at the peak $k=2k_G$: $5\times10^{51}$ for $m=125$~GeV (the Higgs mass, anticipating Sec.~\ref{sec:couplings}). The quartic dominates, $\eta_k>1$, between $k_*=4\times10^{-7}k_G$ and $k_e=11.7k_G$. Averaged over the band with the mode measure $k^2\dd k$, $\eta_k$ is $5\times10^{49}$. The quartic modes are indeed stiff.

The slope $\frac{\dd\epsilon_1}{\dd k}$ is the group velocity of the lowest excitation. It is about $2\times10^{41}c$ at $k_G$ and still $7\times10^{30}c$ at $k_*$, so the dispersion relation alone cannot settle whether signals travel faster than light, and Sec.~\ref{sec:causality} computes the spacelike commutator instead.

\section{CKN matching and the vacuum prescription}\label{sec:matching}
The Hamiltonian of Sec.~\ref{sec:ssb} has two parameters, the coupling $\zeta$ and the envelope scale $k_G$. We fix them with two conditions. The first sets the peak of the ground-state energy per mode (the peak in the two panels of Fig.~\ref{fig:E0}) to the Planck energy (Sec.~\ref{sec:excess}). The second requires the vacuum energy of the scalar in a region of size $L$ to saturate the Cohen-Kaplan-Nelson bound. That bound states that the energy in a region cannot exceed the mass of a black hole of the same size, since otherwise the region would have already collapsed. For a sphere of radius $L$, with volume $\frac{4\pi L^3}{3}$ and Schwarzschild mass $\frac{L}{2G_N}=\frac{L M_{Pl}^2}{2}$,
\begin{equation}
 \frac{4\pi L^3}{3}\,\rho\leq\frac{L\Mpl^2}{2}\qquad
 \rho_{CKN}(L)=\frac{3\Mpl^2}{8\pi L^2}
 \label{eq:rhoCKN}
\end{equation}
The bound is an inequality; we take it to be saturated, as a principle, by the total vacuum energy of the region, $\rho_{vac}=\rho_{CKN}$. See Sec.~\ref{sec:gdep} for an alternative matching condition. Before imposing saturation, we must clarify which energies count toward this density.
 
\subsection{Depletion vs. normal ordering}\label{sec:prescription}
Summed over all modes, the free zero-point energy $\frac{\Omega_k}{2}$ of the scalar is far larger than $\rho_{CKN}$. The modes below $125$~GeV alone give about $10^{53}\rho_{CKN}$. In flat-space quantum field theory this energy is removed by normal ordering. The Hamiltonian is defined with the free zero-point energy subtracted, mode by mode, for bosons and fermions alike (for fermions the subtracted term is negative), which is permitted because without gravity only energy differences are observable~\cite{Jerome}. We do the same. Normal ordering does not determine what gravitates, this is therefore, an assumption. That is the usual form of the cosmological-constant problem and it is not addressed here.

Normal ordering does not remove the change of the ground energy produced by an interaction. For each Fourier mode the quartic shifts the ground energy by
\begin{equation}
 \delta\mathcal E_0(k)=\mathcal E_0(k)-\frac{\Omega_k}{2}\geq0
 \label{eq:shift}
\end{equation}
Normal ordering subtracts the free zero-point sum of empty space, a reference held fixed whatever boundary or interaction is present, so it leaves the Casimir energy, which is the difference between the ground energies with and without the plates, exactly as it leaves the shift of Eqn.~\eqref{eq:shift}.
The shift is non-negative because $U_4$ is a non-negative operator. We take it to gravitate in full, as assumed in Sec.~\ref{sec:intro}; its density $\rho_Q$ is the quartic contribution to the total vacuum energy. The stiff band lies between the two roots $k_*$ and $k_e$ of $\eta_k=1$. Its mode density follows from the continuum phase-space measure:
\begin{align}
 n&=\int_{k_*<|\vec k|<k_e}\frac{\dd^3k}{(2\pi)^3}
   =\frac1{2\pi^2}\int_{k_*}^{k_e}k^2\,\dd k\nonumber\\
  &=\frac{k_e^3-k_*^3}{6\pi^2}\simeq\frac{k_e^3}{6\pi^2}
 \label{eq:modedensity}
\end{align}
Here $n$ counts the modes whose response is strongly changed by the quartic. The lower edge is negligible in this count. No modes outside the band are removed.

An alternative approach, due to Banks and Draper~\cite{CKN2,CKN3}, depletes the single-particle density of states of every field by a factor $D_L(E)=\min\big[1,\big(\frac{\Mpl}{E^2L}\big)^3\big]$ and thereby reduces the zero-point energy sum itself to a value of order $\rho_{CKN}$. We do not use it for this massive scalar for the following reasons
\begin{enumerate}
\item Their counting rests on the free-field density of states, $\propto E^2$, whose zero-point energy grows as $E^4$ per logarithmic interval; once the massive species are included in the count, as the collapse bound requires, the density of states is reduced too far, as the next item quantifies, and 
\item when we use the density of states to, for instance, count the number of electron states, this implementation is excluded by the number of electrons in the universe (\ref{app:BD}).
\end{enumerate}
In addition, its authors leave the microscopic implementation open. The density-of-states prescription alone also does not specify the treatment of the condensate contributions, the Higgs potential at its minimum and the QCD condensate, which are not sums over single-particle states; we retain them in $\rho_{const}$ below. Regardless, it is possible that a reformulated depletion scheme that is sensitive to particle masses and vacuum energy terms might reduce the effect of such terms.

Hence, applying the normal ordering prescription, the CKN bound is applied to the vacuum energy. The bound constrains the energy in the region: a configuration whose total energy exceeds $\Mpl^2R$ is a black hole, not a field-theory state. It does not remove or deplete any modes, and so it does nothing to the zero-point energy of the modes (which is why normal ordering is needed). For the ground state of the region it reads $\rho_{vac}R^3\leq\Mpl^2R$, with $\rho_{vac}$ the total vacuum energy density, and since $\rho_{vac}$ is uniform it holds in every sub-region once it holds at the largest scale $L$. We take the bound to be saturated by the total, as a principle:
\begin{equation}
 \rho_{vac}=\rho_Q+\rho_{const}=\rho_{CKN}(L)
 \label{eq:law}
\end{equation}
Here $\rho_{const}$ consists of every constant contribution, i.e.,
\begin{enumerate}
\item The classical energy of the Higgs potential at its minimum, which is $-\frac{\lambda_Hv^4}{4}\simeq-10^8$~GeV$^4$ for the unshifted potential
\item The QCD condensate, of order $10^{-3}$~GeV$^4$
\item The loop contributions of the other interactions not removed by the stated subtraction
\end{enumerate}
These contributions can be enormously larger than $\rho_{CKN}$, and the Higgs and QCD contributions are negative in the conventions used here. Requiring contributions of electroweak size to sum to the observed density involves a cancellation of order one part in $10^{55}$. Under the exact saturation principle of Eqn.~\eqref{eq:law}, the conditions on the retained constants instead take the following form:
\begin{enumerate}
\item First, under the postulated exact saturation principle the total is fixed by CKN saturation: $\rho_Q$ is positive and adjustable, and the principle fixes it to whatever brings the total to $\rho_{CKN}$.
\item Hence a vacuum energy sum for $\rho_{const}$ does not need to track to zero.
\item Second, the gap is filled by the one term that tracks the horizon, i.e., $\rho_Q$. The condition on the other constant terms from the principle alone is therefore one-sided, $\rho_{const}\in(-\infty,\rho_{CKN}]$. Explicitly, the two terms in Eqn.~\eqref{eq:law} sum to $\rho_{CKN}$ exactly. Since $\rho_Q\geq0$, the restriction falls on the residual alone, $\rho_{const}\leq\rho_{CKN}$, with $\rho_Q=\rho_{CKN}-\rho_{const}$. A negative sum for $\rho_{const}$ is compensated by $\rho_Q$, which then exceeds $\rho_{CKN}$ and enlarges the band, $k_e\propto\rho_Q^{1/3}$. A large negative constant balanced by a large positive nonlocal term is allowed.
\item A more negative $\rho_{const}$ requires a larger positive $\rho_Q$ to keep their sum equal to $\rho_{CKN}$. This broadens the band and makes the predicted extra force stronger at shorter distances (Sec.~\ref{sec:couplings}). In the point-source approximation, the force envelope reaches about $10^{-3}$ of gravity at a range of $20$~cm when $\rho_Q\sim10^{17}\rho_{CKN}$, corresponding to $\rho_{const}\sim-(50~\mathrm{eV})^4$. This is a scale for comparison with experiments, not an experimental bound: the oscillating force must be integrated over the actual source and test-body geometry. If that analysis permitted a residual as negative as $-(50~\mathrm{eV})^4$, the allowed interval would be about one part in $10^{38}$ of the electroweak scale $(100~\mathrm{GeV})^4$ (this tuning is $\frac{(50 eV)^4}{10^8 GeV^4}$), rather than one part in $10^{55}$. The saturation principle assigns the compensating value to $\rho_Q$; the microscopic dynamics that would select that value remains to be determined.
\end{enumerate}
In what follows, however, our benchmark is computed for $\rho_{const}=0$. 

Under normal ordering no free field contributes to the vacuum energy, massive or massless, so the matching involves the scalar alone. Nonlocal interactions of the same kind in the other fields, if present, would add their own shifts to Eqn.~\eqref{eq:law} and are not considered here.

\subsection{Matching of the interaction energy}\label{sec:excess}
With the free zero-point energy normal-ordered, Eqn.~\eqref{eq:law} fixes the shift of Eqn.~\eqref{eq:shift} integrated over wavevectors,
\begin{equation}
 \rho_Q=\frac1{2\pi^2}\int_0^\infty k^2\,\delta\mathcal E_0(k)\,\dd k=\rho_{CKN}-\rho_{const}
 \label{eq:combined}
\end{equation}
We set the benchmark $\rho_Q$ by assuming $\rho_{const}=0$ and hence $\rho_Q=\rho_{CKN}$. A negative residual enlarges the band, rescaling $k_G$ by $\big(1-\frac{\rho_{const}}{\rho_{CKN}}\big)^{1/3}$ and $\zeta$ accordingly (Sec.~\ref{sec:prescription}). The powers of $L$ below hold for $|\rho_{const}|\ll\rho_{CKN}$. A fixed residual comparable to $\rho_{CKN}$ enters through $\rho_{CKN}(L)-\rho_{const}$, giving $\frac{\dd\ln k_G}{\dd\ln L}=-\frac23\frac{\rho_{CKN}}{\rho_{CKN}-\rho_{const}}$: a dominant negative constant leaves $k_G$ nearly independent of $L$, with $\zeta\propto L^{-3}$ from the peak normalization, while the total still obeys Eqn.~\eqref{eq:law}.
The conditions we set before fix $\zeta$ and $k_G$. The first sets the peak of the ground energy to the Planck energy. In the pure-quartic form of Eqn.~\eqref{eq:E0} the peak lies at $k=2k_G$, and $\mathcal E_Q(2k_G)=\Mpl$ gives
\begin{align}
 \zeta&=\frac{e^4\Mpl^3}{\gamma_4^3L^3(2k_G)^8}
 \label{eq:FG}
\end{align}
The second is Eqn.~\eqref{eq:combined}. With $\mathcal E_Q$ of Eqn.~\eqref{eq:E0} in place of $\delta\mathcal E_0$ (the difference being minuscule), we get $\rho_Q=\frac{\Mpl k_G^3I_G}{2\pi^2}$ and
\begin{equation}
 k_G^3=\frac{3\pi}{4I_G}\frac{\Mpl}{L^2}\qquad k_G\propto\Mpl^{1/3}L^{-2/3}\qquad \zeta\propto\Mpl^{1/3}L^{7/3}
 \label{eq:kG}
\end{equation}
where $I_G=\int_0^\infty x^2F_G(x)\,\dd x=11.58$, with $F_G(x)=\frac{\mathcal E_Q(xk_G)}{\Mpl}=\big(\frac x2\big)^{8/3}e^{(4-x^2)/3}$. The powers of $\Mpl$ and $L$ in Eqn.~\eqref{eq:kG} follow from dimensional counting alone: once the peak is normalized to $\Mpl$, $\mathcal E_Q(k)=\Mpl F_G(k/k_G)$ with $F_G$ dimensionless, so $\rho_Q\propto\Mpl k_G^3$ whatever the function $F_G$, and the exponent $\frac{8}{3}$ enters only through the number $I_G$.

This is the UV/IR relation of Eqn.~\eqref{eq:mainclosureintro}, {\it {i.e.,}} $k_G\simeq0.17\,k_{Pl}^{1/3}k_{box}^{2/3}$. We denote the present benchmark size by $L_U$. Under the event-horizon identification of Sec.~\ref{sec:hde}, $L_U=R_h(t_0)$, where $t_0$ is the present cosmic time. For $m=125$~GeV, $\Mpl=1.2209\times10^{19}$~GeV and the observed dark-energy density $2.5\times10^{-47}$~GeV$^4$,
\begin{equation}
\begin{aligned}
 L_U&\simeq1.665\times10^{26}\,\mathrm m\\
 k_{G,U}&\simeq1.517\times10^{-22}\,\GeV\\
 \zeta_U&\simeq1.145\times10^{106}\,\GeV^{-2}\\
 g\overline C_{L_U}^{\,4}&\simeq1.609\times10^{22}
\end{aligned}
\label{eq:benchmark}
\end{equation}
The peak lies at $2k_G=3.03\times10^{-22}$~GeV, a wavelength of $4100$~km, and the band ends at $k_e=11.71k_G$. Since the density is an input and $\zeta$ is fitted, these numbers are calibrated. The matching is made once, at the largest scale $L$. The ground-state density is uniform, so a region of size $R<L$ inside it holds $\rho_{CKN}(L)R^3<\Mpl^2R$ and obeys the bound without saturating it. Nested regions cannot all be saturated by a uniform density, and $k_G$ and $k_e$ are properties of the Hamiltonian of the whole region.

\section{Homogeneous cosmology under the horizon prescription}\label{sec:hde}
Holographic dark energy~\cite{Li2004} identifies $L$ with the future event horizon $R_h$, the largest distance from which a light signal sent now can still reach us. With $L=R_h$ in Eqn.~\eqref{eq:combined}, the vacuum energy follows the horizon,
\begin{equation}
 R_h(t)=a(t)\int_t^\infty\frac{\dd t'}{a(t')}\qquad
 \rho_{vac}=\rho_{CKN}(R_h)=\frac{3\Mpl^2}{8\pi R_h^2}
 \label{eq:HDE}
\end{equation}
Everything in this section follows from Eqn.~\eqref{eq:HDE} alone and is the standard holographic fluid. If the component is separately conserved, $\rho_{vac}\propto R_h^{-2}$ and $\dot R_h=HR_h-1$ fix its equation of state,
\begin{equation}
 w=-\frac13-\frac23\sqrt{\Omega_{vac}}
 \qquad \Omega_{vac}=\frac{8\pi\rho_{vac}}{3\Mpl^2H^2}=\frac{1}{H^2R_h^2}
 \label{eq:w}
\end{equation}
For $\Omega_{vac,0}=0.69$, $w_0\simeq-0.89$, and $w\to-1$ as $\Omega_{vac}\to1$. Integrating $\frac{\dd\Omega_{vac}}{\dd\ln a}=\Omega_{vac}(1-\Omega_{vac})(1+2\sqrt{\Omega_{vac}})$ with pressureless matter and no radiation gives Table~\ref{tab:history}.

\begin{table}[t]
\centering
\caption{Conditional homogeneous history under the holographic-fluid prescription, for $\Omega_{vac,0}=0.69$ and $\Omega_{m,0}=0.31$, with radiation neglected. These values are not a fit to cosmological observations.}
\label{tab:history}
\begin{tabular}{lrrrrrr}
\toprule
$z$ & $0$ & $0.2$ & $0.5$ & $1$ & $2$ & $3$\\
\midrule
$w$ & $-0.89$ & $-0.84$ & $-0.78$ & $-0.70$ & $-0.60$ & $-0.55$\\
$\Omega_{vac}$ & $0.69$ & $0.58$ & $0.45$ & $0.30$ & $0.16$ & $0.11$\\
\bottomrule
\end{tabular}
\end{table}

This calculation has three components.
\begin{enumerate}
\item We posit the identification $L=R_h$.
\item We assume that as the universe expands, the energy in the quartic modes is conserved separately from matter and radiation. The impact of interactions with other matter has not been analyzed, as has the  evolution of the couplings discussed below.
\item As the universe expands, it is a reasonable extension of CKN to ask that the bound remain saturated at all times. That requires, from Eqn.~\eqref{eq:kG} and for $|\rho_{const}|\ll\rho_{CKN}$, that $\zeta\propto R_h^{7/3}$ and $k_G\propto R_h^{-2/3}$, with the peak of the ground energy held at $\Mpl$. We require this only for epochs after the electroweak and QCD transitions. Before and through them the interaction need not have this form, and more work is needed.
\end{enumerate}
With these, the total density obeys $\rho_{vac}=\rho_{CKN}(R_h)\propto R_h^{-2}$ at every epoch, with $\rho_Q\simeq\rho_{vac}$ when $\rho_{const}$ is negligible. The total component is exactly the holographic fluid of Eqn.~\eqref{eq:HDE}, so Eqn.~\eqref{eq:w} and Table~\ref{tab:history} follow with no further input. Since $\rho_{const}$ in Eqn.~\eqref{eq:law} is fixed while $\rho_{CKN}(R_h)$ falls as the horizon grows, the principle can be maintained with $\rho_Q\geq0$ only if the constants sum to no more than the asymptotic value of $\rho_{CKN}(R_h)$, which the fluid approaches as $\Omega_{vac}\to1$, not merely below the present one. The band itself moves: its edge $k_e=11.7k_G$ and the mode density $n=\frac{k_e^3}{6\pi^2}\propto R_h^{-2}$ fall as the horizon grows, so the modes that carry the vacuum energy are not a fixed set, and which modes enter and leave the band, and how $\zeta$ tracks $R_h$, which depends on the future expansion, depend on detailed analysis with a dynamical clock and metric.

The resulting fluid is disfavored by some current data combinations. In the notation of the holographic literature, $\rho_{vac}=\frac{3c^2\Mpl^2}{8\pi R_h^2}$, saturation is $c=1$: late-time data allow it, cosmic-microwave-background data prefer $c<1$, and Bayesian comparisons disfavor the fluid relative to $\Lambda$CDM, strongly once those data are included~\cite{LiHDE2025}.

\section{Long-distance signalling}\label{sec:causality}
The interaction is instantaneous in the preferred frame, so a local source might influence a distant region before light arrives. However, as we will show below, the stiffness of the quartic band makes the signal exceedingly small.  Consider the vacuum commutator. With $\vec r=\vec x-\vec x'$, Eqn.~\eqref{eq:pairfield} gives
\begin{align}
 \Delta(\vec r,t)&\equiv i\avg{[\phi(\vec x,t),\phi(\vec x',0)]}\nonumber\\
 &=\frac1V\sum_{\vec k}\cos(\vec k\cdot\vec r)\,c_k(t)\nonumber\\
 c_k(t)&=i\avg{[x_k(t),x_k(0)]}
 \label{eq:Delta}
\end{align}
because different pairs commute, parity, which sends $y_k$ to $-y_k$, removes the $x_k$-$y_k$ cross terms, and the rotation symmetry of Eqn.~\eqref{eq:pair} makes the $x_k$ and $y_k$ responses equal. Here $c_k$ is the motion of mode $\vec k$ after a unit impulse. Inserting the eigenstates of Eqn.~\eqref{eq:pair}, $c_k=2\sum_n|\langle n|x_k|0\rangle|^2\sin\omega_{n,k}t$, so $|c_k|\leq\avg{S}_{0,k}$.

For a free pair $c_k=\frac{\sin(\Omega_kt)}{\Omega_k}$, and in the continuum
\begin{equation}
 \Delta_0(r,t)=\frac1{2\pi^2r}\int_0^\infty k\sin(kr)\,\frac{\sin\Omega_kt}{\Omega_k}\,\dd k=0\qquad\text{for }r>|t|
 \label{eq:Delta0}
\end{equation}
This follows by writing $k\sin kr=-\partial_r\cos kr$ and using the rapidity, $k=m\sinh u$, with $r=s\cosh\alpha$ and $t=s\sinh\alpha$, which makes the integrand odd. Individual terms need not vanish, so the result is a property of the complete sum with $\Omega_k=\sqrt{k^2+m^2}$. To consider the effect of just the quartic term for spacelike separation, we compute
\begin{equation}
 \Delta(\vec r,t)=\frac1V\sum_{\vec k}\cos(\vec k\cdot\vec r)\Big[c_k(t)-\frac{\sin\Omega_kt}{\Omega_k}\Big]\qquad r>|t|
 \label{eq:Dk}
\end{equation}
so only modes whose response the quartic changes contribute.

The stiff quartic band is the range $\eta_k>1$. With $\lambda_k$ of Eqn.~\eqref{eq:pair} it runs from the lower edge $k_*$, where $\lambda_k=m^3$, to the upper edge $k_e$, where the Gaussian has brought $\lambda_k$ back to $m^3$:
\begin{align}
 k_*&\simeq6\times10^{-29}~\GeV\nonumber\\
 k_e&=11.709\,k_G=1.78\times10^{-21}~\GeV
 \label{eq:edges}
\end{align}
at the benchmark, with wavelengths $\frac{2\pi}{k}$ of $129$~AU and $698$~km. In the above band the first term in the bracket of Eqn.~\eqref{eq:Dk} is negligible: the quartic modes respond at a fraction of order $\eta_k^{-1/3}$ of the free response, $10^{-17}$ at the peak, so we set $c_k=0$ for $k_*<k<k_e$. This is the only contribution to the signal. With $\Omega_k\simeq m$ in the band, since $k_e\ll m$, and the lower edge neglected, $\big(\frac{k_*}{k_e}\big)^3\simeq5\times10^{-23}$, the sum over $k<k_e$ is $-\frac{\sin mt}{m}$ times $\frac1V\sum_{|\vec k|<k_e}\cos(\vec k\cdot\vec r)$, the Fourier transform of the band, which gives
\begin{equation}
 \Delta(r,t)\simeq-\frac{\sin mt}m\,\frac{\sin k_er-k_er\cos k_er}{2\pi^2r^3}\simeq-n\,\frac{\sin mt}m
 \label{eq:Deltaband}
\end{equation}
For $r\ll\frac{1}{k_e}=111$~km, expanding the functions gives the last expression in Eqn.~\eqref{eq:Deltaband}. Equation~\eqref{eq:modedensity} gives $n\simeq9.5\times10^{-65}$~GeV$^3$ at the benchmark, approximately one mode per $(433~\mathrm{km})^3$.

To estimate how important this term is, consider the following ratio. The natural comparison is the free commutator just inside the light cone (where its magnitude is biggest), which for a massive field tends to $-\frac{m^2}{8\pi}$. Eqn.~\eqref{eq:Deltaband} is bounded by $\frac nm$, so at the benchmark
\begin{equation}
 \frac{8\pi|\Delta|}{m^2}\leq\frac{8\pi n}{m^3}=\frac4{3\pi}\Big(\frac{k_e}m\Big)^3=1.2\times10^{-69}
 \label{eq:Dratio}
\end{equation}
The spacelike commutator exists only because of the quartic interaction term. It is small for two reasons that combine into $\big(\frac{k_e}{m}\big)^3$, namely
\begin{enumerate}
\item The Gaussian confines the band to wavelengths above $700$~km, so $n$ is small, 
\item The mass term limits the response each band mode would have had to $\frac1m$. Without the mass the free response of a band mode would be $\frac{\sin(kt)}{k}$, which grows without bound as $k\to0$ and $t\to\infty$, and the spacelike commutator would be comparable to the field's own fluctuations at separations near $\frac{1}{k_e}$.
\end{enumerate}

We might consider the following questions about Eqn.~\eqref{eq:Deltaband}: how accurate it is, for how long it holds, and what it would mean for an experiment.

\emph{Accuracy.} Setting the quartic mode response to zero is asymptotically accurate deep in the quartic-dominant region. The modes near its edge still respond a little, and including their exact response changes $n$ in Eqn.~\eqref{eq:Deltaband} by a few percent.

\emph{Duration.} Modes just outside the band have frequencies slightly above the free ones, so with time they fall out of step with the free response they are compared with in Eqn.~\eqref{eq:Dk}. The further a mode lies outside the band, the smaller its frequency shift and the longer this takes. The band therefore appears to widen slowly with time, and since the quartic falls off as a Gaussian beyond $k_e$, $n$ grows only logarithmically, by about a factor of two over the age of the universe.

\emph{Experiment.} Switch on a point source $J(t,\vec x)=J_0\Theta(t)\delta^{(3)}(\vec x)$ at $t=0$, with interaction Hamiltonian $-\int\dd^3x\,J\phi$. Integrating Eqn.~\eqref{eq:Deltaband} over the source history gives the missing-band contribution before light arrives, for $0<t<r$,
\begin{align}
 \delta\phi_{band}(r,t)&\simeq
 -\frac{J_0}{m^2}\mathcal B(r)\,[1-\cos(mt)]
\nonumber\\
 \mathcal B(r)&=\frac{\sin(k_er)-k_er\cos(k_er)}{2\pi^2r^3}
 \label{eq:switchedsource}
\end{align}
The ordinary free modes in this band would contribute the static field $+J_0\mathcal B(r)/m^2$. The minus sign in Eqn.~\eqref{eq:switchedsource} means that the stiff band removes that contribution; it is opposite to the free band's share, not a reversal of the entire Yukawa field. The sudden switch produces oscillations between zero and twice this static correction. For $r\ll1/k_e$, $\mathcal B(r)\simeq n$. The potential correction from removing the same band is Eqn.~\eqref{eq:dV}. For Higgs-coupled nucleons, Sec.~\ref{sec:couplings} gives a maximum static-force correction of a few times $10^{-15}$ of gravity. A signalling test would have to resolve the time-dependent change before the light-travel time, which is about $0.4$~ms across $111$~km.

Equation~\eqref{eq:Deltaband} is the linear response of the vacuum to weak sources. Strong or macroscopic coherent sources, for which the band's stiff response is no longer linear, are not covered, and for those, exact microcausality is not established, since the commutator is not precisely zero.

\section{Dependence on the coupling}\label{sec:gdep}

There is a different way to establish and use the matching conditions. If we simply assumed that $g$ in Eqn.~\eqref{eq:S4} were of order one, with $\Cbar=1$ so that $\zeta=gL^2$ in place of the $g\Cbar^{\,4}=1.6\times10^{22}$ of Eqn.~\eqref{eq:benchmark}, a naturalness argument, and only applied matching with the CKN condition, we would find that the dispersion peak sits at $1.5\times10^{15}$~GeV (rather than the Planck scale), the band ends at $6$~km rather than $111$~km, the commutator ratio is $7\times10^{-66}$, the force envelope $8\times10^{-13}$ of gravity, the powers in Eqn.~\eqref{eq:kG} become $\frac{6}{17}$ and $\frac{11}{17}$, and the equation of state of Sec.~\ref{sec:hde} is unchanged. The reason is that both choices impose the same total density, $\rho_{vac}=\rho_{CKN}(R_h)\propto R_h^{-2}$, and use the same conservation law. Changing the coupling prescription changes how the scalar spectrum supplies that density, but not how the total density depends on the horizon. Since the equation of state in Eqn.~\eqref{eq:w} follows from that horizon dependence, it is unchanged.

\section{Higgs and gravitational extensions}\label{sec:couplings}
\subsection{A gauge-invariant effective interaction}
We identify the scalar with the radial Higgs mode after symmetry breaking ($\langle\HH\rangle\neq0$). The scalar of Sec.~\ref{sec:ssb} is a single real field, whereas the Higgs is a doublet $\HH$ on which the electroweak group acts, so $U_4$ cannot be written in terms of one of its components without choosing a gauge. The length of the doublet is the combination that is gauge invariant and non-negative, hence we build $U_4$ from it. The doublet keeps its gauge and Yukawa interactions and its potential $\lambda_H\big(\HH^\dagger\HH-\frac{v^2}{2}\big)^2$; the only addition is $U_{4,h}$. We define
\begin{equation}
 h=\sqrt{2\HH^\dagger\HH}-v\qquad
 U_{4,h}=\frac\zeta V\sum_{\vec k}k^8e^{-(k/k_G)^2}|\widetilde h_{\vec k}|^4\geq0
 \label{eq:Hradial}
\end{equation}
The coupling $g$ of the spacetime-nonlocal form of Eqn.~\eqref{eq:S4}, with $h$ in place of $\phi$, is dimensionless, so the interaction is marginal. Equation~\eqref{eq:Hradial} is a proposed effective operator.

\subsection{Tests for the new term}
\paragraph{Static force test} A scalar coupled to matter produces a force between two bodies. For an ordinary massive field this is the Yukawa force, which falls off exponentially beyond the distance $\frac1m$; for the Higgs that distance is $10^{-18}$~m, irrelevant for experiments. However, the Yukawa force is a sum over all the modes of the field, each contributing in proportion to how far it is displaced by a static source, and the exponential fall-off at large distance is the result of a near-perfect cancellation among those contributions. In this model the modes of the band are stiff and are hardly displaced at all, so their share of the sum (over wavevectors) is missing. Hence, the above cancellation fails. What is left is a weak force between macroscopic bodies with a range set by the band's wavelength, about $111$~km, rather than by $\frac1m$. We estimate this force by subtracting the band's share from the Yukawa sum, and compare the result with the two kinds of experiment that bound new forces: inverse-square tests, which constrain any extra force whatever the bodies are made of, and equivalence-principle tests, which constrain the part that depends on composition.

Two static sources with scalar charges $g_A$ and $g_B$ feel the Yukawa potential expressed as a sum over modes,
\begin{equation}
 V_\phi(r)=-\frac{g_Ag_B}{2\pi^2r}\int_0^\infty\frac{k\sin(kr)}{k^2+m^2}\,\dd k = -\frac{g_Ag_B}{4\pi r} e^{-mr}
 \label{eq:staticpotential}
\end{equation}
in which $\frac{1}{k^2+m^2}$ is the free propagator. A stiff quartic mode has a susceptibility smaller by a factor of order $\eta_k^{-2/3}$ deep in the band, and $0.32$ of the free value at the edge, where $\eta_k=1$. Neglecting the band's response altogether gives an estimate up to the edge, by removing that bit entirely. Removing it from the above equation, with $k^2+m^2\simeq m^2$ since $k_e\ll m$, leaves
\begin{align}
 \delta V(r)&=\frac{g_Ag_B}{2\pi^2m^2r}\int_0^{k_e}k\sin(kr)\,\dd k\nonumber\\
 &=g_Ag_B\,\frac{\sin k_er-k_er\cos k_er}{2\pi^2m^2r^3}
 \label{eq:dV}
\end{align}
a correction to the potential that is a positive constant for $r\ll\frac{1}{k_e}=111$~km, falls through zero at about $500$~km and oscillates about zero beyond, so that the force it produces is negligible at short distance, repulsive out to $640$~km, and alternating in sign thereafter with period $\frac{2\pi}{k_e}=698$~km.

We can compute the correction for nucleons.
For a body $A$ (mass $m_A$) the charge is $g_A=\frac{\partial m_A}{\partial v}$, since the Higgs couples to matter through the Yukawa term. For a nucleon the dependence on $v$ enters through the quark masses $m_q=\frac{y_qv}{\sqrt2}$, so $g_N=\frac{m_N}{v}\sum_qf_q^N$ with $f_q^N=\frac{m_q}{m_N}\frac{\partial m_N}{\partial m_q}$ the nucleon sigma terms; the light quarks contribute about $0.1$ and the heavy quarks, through the gluon condensate, $\frac{2}{27}$ each~\cite{SVZ}, for a total $f_N\simeq0.3$, so that $g_N=\frac{f_Nm_N}{v}\simeq1.1\times10^{-3}$. The force from Eqn.~\eqref{eq:dV} is the gradient of $\delta V$, and its ratio to the Newtonian force between the same two nucleons, $\frac{m_N^2}{\Mpl^2r^2}$, depends on the separation. It vanishes at short distance, where $\delta V$ is constant, and is largest at about $430$~km, where it equals $2.5\times10^{-15}$.

The charge is approximately proportional to mass, hence the correction is composition independent at least to leading order. The existing inverse-square tests, which allow a new force of $10^{-3}$ to $10^{-8}$ of gravity at distances from $1$~m to $10^5$~km~\cite{Will,Adelberger}, would not be sensitive to this correction. The composition-dependent part is set by the variation of $\frac{g_A}{m_A}$ between materials. For a scalar coupled through the quark and electron masses this variation is of order $10^{-3}$ between test-body materials~\cite{DamourDonoghue}, which puts the composition-dependent force near $10^{-18}$ of gravity, three orders of magnitude below the equivalence-principle bound of Ref.~\cite{MICROSCOPE}.

\begin{figure}[t]
\centering
\includegraphics[trim=0 16bp 0 0,clip,width=0.7\columnwidth]{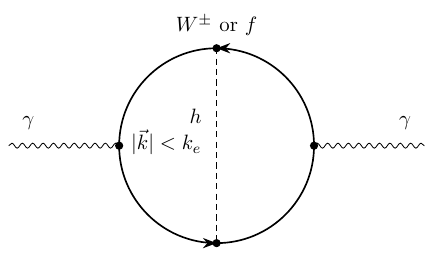}
\caption{Photon self-energy diagram with a Higgs exchanged across a charged loop.}
\label{fig:twoloop}
\end{figure}

\paragraph{Radiative coefficients} The reduced scalar Hamiltonian is diagonal in $\vec k$, so its conjugate pairs do not scatter into one another. The quartic changes the scalar propagator. Above the quartic region the Gaussian makes the propagator progressively approach its free form.

If the scalar were the radial Higgs, its gauge and Yukawa couplings would connect it to other fields. The preferred-frame dependence can then enter their kinetic terms through radiative corrections~\cite{Abel2006,Collins}. In the preferred frame the photon operator $n^\rho n^\nu F_{\rho\mu}F^\mu{}_\nu$ is $E^2$. The isotropic coefficient $\tilde\kappa_{tr}$ measures the change in the electric kinetic term relative to the magnetic one and hence affects photon propagation. For standard charged-particle dispersion, the quoted bound is $-9\times10^{-16}<\tilde\kappa_{tr}<6\times10^{-20}$~\cite{KlinkhamerSchreck}.

At zero frequency and the band peak $k=2k_G$, the ordinary Euclidean Higgs propagator $D_H^E$ in Fig.~\ref{fig:twoloop} is approximately $1/m_H^2$, where $m_H$ is the Higgs mass. With the quartic interaction, the lowest excitation gap of the same mode is of order $\Mpl$ (Fig.~\ref{fig:E0}), and its static propagator $D_Q^E$ is of order $1/\Mpl^2$. The ratio of the two Higgs propagators for this mode is
\begin{equation}
 \frac{D_Q^E(0,2k_G)}{D_H^E(0,2k_G)}
 \sim \left(\frac{m_H}{\Mpl}\right)^2
 \label{eq:radiativeestimate}
\end{equation}
Now sum each static propagator over $\dd^3k/(2\pi)^3$ up to the same cutoff $\Lambda$. The band contributes $B_H\simeq n/m_H^2$ for the ordinary Higgs. For the quartic Higgs, $B_Q$ is the integral of $D_Q^E(0,k)$ over those same modes. Write $A(\Lambda)$ for the common contribution from the remaining momenta, neglecting the small change outside the band. The ratio of the quartic integral $N$ to the ordinary integral $D$ is then
\begin{equation}
 \frac{N}{D}\simeq\frac{A+B_Q}{A+B_H}
 =1-\frac{B_H-B_Q}{A+B_H}
 \label{eq:staticintegralratio}
\end{equation}
For $\Lambda\gg m_H$, $A$ grows linearly with $\Lambda$ because $k^2/(k^2+m_H^2)$ approaches one. The difference of the two static integrals is finite even with the full Gaussian, so their ratio is close to one - the difference is of order $\frac{k_e^3}{m_H^2 \Lambda} \sim 10^{-69}$ (for $\Lambda \sim m_H$), well under experimental limits for the photon coefficient $\tilde\kappa_{tr}$.

\subsection{Coupling to gravity}
The flat-space Hamiltonian does not by itself determine how $\rho_Q$ sources gravity. A possible extension makes the metric and clock dynamical,
\begin{equation}
 S=S_{EH}[g]+S_T[g,T]+S_{SM}[g,\psi]-\int\dd\tau\,U_{4,\Sigma_\tau}[\alpha,h;g]
 \label{eq:gravity}
\end{equation}
where $S_{EH}$ is the Einstein-Hilbert action, $S_T$ is the clock action, and $U_4$ acts on slices $\Sigma_\tau$ of constant $T$. Consistency requires
\begin{equation}
 \nabla_\mu\big(T^{\mu\nu}_{SM}+T^{\mu\nu}_{T}+T^{\mu\nu}_{U_4}\big)=0
 \label{eq:conservation}
\end{equation}
After the same free-vacuum subtraction used above, the scalar stress must reproduce $\rho_Q$, including the changed expectations of both the free Hamiltonian and $U_4$.

For a canonical free clock, write $T=\bar T(t)+\chi(t,\vec x)$, where the homogeneous solution obeys $\dot{\bar T}=M^2a^{-3}$, with $a=1$ today. Initial conditions fix the present clock rate $M^2$, and its homogeneous energy density is $\rho_T=M^4/(2a^6)$. If this free evolution extends back to nucleosynthesis, its energy-density bound requires $M\lesssim$ a few neV. Locally today the background is approximately linear in time. The fluctuation has two effects: $\vec\theta=\nabla\chi/M^2$ tilts the clock slices, while $\epsilon=\partial_t\chi/M^2$ changes their rate relative to the background time. Both must be small, in our approximation (we keep upto quadratic order in fluctuations). For free-clock vacuum fluctuations coarse-grained at wavenumber $q$, their root-mean-square sizes are of order $q^2/(4\pi M^2)$, suggesting $q\ll M$. Covering the quartic band then requires $k_e\simeq2\times10^{-12}$~eV $\ll M\lesssim$ a few neV. At other epochs the corresponding local clock scale is set by ${\dot{\bar T}}=M^2a^{-3}$.

In ~\ref{app:clock} we model $\chi$ as a single sinusoidal
ripple of wavenumber $q$ with a time-dependent amplitude. We keep
terms through second order in the perturbation. The lapse couples to the scalar energy, including the quartic contribution, and the shift couples to the scalar momentum. Equation~\eqref{eq:clockdispersion} gives the resulting low-frequency dispersion relation. The static coefficient $K_0(q)$ reduces the clock's spatial restoring force, while the temporal coefficient $D(q)\ge0$ increases its inertia. For $0<K_0<M^4$ the mode is stable and its phase speed is subluminal; evaluating $D(q)$ determines the size of the additional slowing.

A completion must still determine the coupled matter-clock retarded response, the stress associated with horizon-dependent couplings, and the gravitational collapse relation underlying CKN. Its gravitational-wave and preferred-frame coefficients must also satisfy observational bounds~\cite{GW170817,SME,Will}. The clock calculation therefore supplies a conditional consistency check; it does not establish a gravitational derivation of CKN saturation - however, the subluminal phase speed of this trial clock mode is compatible with the usual black-hole horizon assumed in our CKN saturation matching argument.

\section{Conclusions}\label{sec:conclusions}

We have presented a program to derive the cosmological constant from quantum field theory. It consists of the following parts
\begin{enumerate}
\item We adopt normal ordering of the Hamiltonian, the usual subtraction prescription.
\item The total vacuum energy includes the Higgs potential minimum, the QCD condensate, loop contributions and possible non-linear field interactions in a preferred Lorentz frame.
\item All of this has to be consistent with the observation of real particles with energies far above the scale of the cosmological constant.
\item A candidate non-linear theory that is still consistent with the notion of high energy states is a momentum-diagonal quartic $\lambda_k=\zeta L^3k^8e^{-(k/k_G)^2}$. It leads to a ground-state energy that rises as $k^{8/3}$ to a peak at $2k_G$ and returns to the ordinary dispersion at high momentum, with a quartic band between $k_*$ and $k_e=11.7k_G$. 
\item The CKN condition is elevated to that of a principle and is used to deduce the free parameters $\zeta$ and $k_G$ in the interaction. The calibration conditions we used, the Planck normalization of the peak and the matching of the interaction energy to the CKN density, fix $k_G\propto\Mpl^{1/3}L^{-2/3}$ by dimensional counting, the Hamiltonian supplying only the coefficient. 
\item As an aside, an alternative and perfectly reasonable calibration uses naturalness to fix the coefficient of the interaction, while CKN is used to fix $k_G$.
\item Superluminal interaction in the preferred frame is suppressed - the spacelike commutator is $\Delta\simeq-\frac{n\sin(mt)}{m}$ at $r\ll\frac{1}{k_e}$, of order $\big(\frac{k_e}{m}\big)^3\sim10^{-69}$ of the free commutator inside the light cone because the quartic region lies far below the mass; the corresponding static force between nucleons is of order $10^{-15}$ of gravity.
\item The constant contributions, i.e., the Higgs potential energy and the QCD condensate and loop contributions, must sum to no more than $\rho_{CKN}$. This is a one-sided condition and leads to less fine-tuning for the cosmological-constant problem compared to its usual form.
\item  With $L$ the event horizon and CKN saturation maintained at every epoch after the electroweak and QCD transitions, the component is the holographic fluid, $w_0\simeq-0.89$.
\item We have written down how such a term would arise for the Higgs sector - a gauge-invariant radial Higgs term best captures the effect discussed in the paper. The quartic suppresses static Higgs exchange at the band peak, but the two static momentum integrals have a ratio close to one at a large common cutoff (Eqn.~\eqref{eq:staticintegralratio}).
\item For an extension to gravity, we have described the simplest possible formulation. We do not expect horizons to be affected and the CKN limit should survive the preferred frame assumption, though this remains to be proven.
\end{enumerate}

\section*{Acknowledgments}
The hospitality of the Rutgers Physics \& Astronomy Department and the NHETC is gratefully acknowledged. Useful discussions with Scott Thomas are also acknowledged.

\section*{Data availability}
No new experimental data were generated. 

\section*{Declaration of generative AI and AI-assisted technologies}
ChatGPT, Claude and Qwen assisted with editing, exposition, algebraic organization and numerical checks during manuscript preparation. The author has written the final paper and is responsible for detailed scientific verification as well as every part of the content.

\appendix

\section{Banks-Draper depletion and the electron count}\label{app:BD}
Refs.~\cite{CKN2,CKN3} implement the CKN bound by depleting the single-particle density of states of every field by $D_L(E)=\min\big[1,\big(\frac{\Mpl}{E^2L}\big)^3\big]$, which brings the zero-point sum of a massless field down to order $\rho_{CKN}$. Read as a literal count of the states of a field of mass $m$, species by species as the vacuum-energy sum requires, the weight gives $\frac{g_sL^3}{2\pi^2}\int k^2D_L(\Omega_k)\,\dd k=\frac{g_s\Mpl^3}{32\pi m^3}$ in any region larger than $\frac{\Mpl}{m^2}$, with $g_s$ the number of spin states. For the electron it is $3\times10^{65}$ states, for the nucleon $4\times10^{55}$, while the observable universe contains $10^{80}$ of each, which is in stark disagreement. Raising the threshold of the same profile until $10^{80}$ electrons fit the states raises the vacuum energy of the massless fields by $\big(\frac{10^{80}}{2.7\times10^{65}}\big)^{2/3}\simeq5\times10^9$. However, the $g$-2 and Lamb-shift tests of Refs.~\cite{CKN2,CKN3} probe loop corrections of order $\big(\frac{1}{mL_{eff}}\big)^p$ and are insensitive to the count. Additionally, the collapse bound constrains the total energy of a configuration - all the species together.

We have experimented with other depletion schemes - a depletion restricted to elementary species heavier than a few GeV passes the count but leaves the zero-point energy of all the light fields in place, $\frac{E_t^4}{16\pi^2}\simeq10^{44}\rho_{CKN}$ below $E_t=1$~GeV alone, and a depletion by energy would act on energetic photons and electrons too. In addition, the depletion prescription does not specify the treatment of the other contributions to the vacuum energy, some of which are large.

\section{The clock's coupling to the quartic modes}\label{app:clock}
Locally the clock is $T=M^2t+\chi$. We take a small trial fluctuation along $x$,
\begin{equation}
 \chi(t,x)=M^2b(t)f(x)\qquad f(x)=\frac{\sqrt2}{q}\sin(qx)
 \label{eq:clocktrial}
\end{equation}
Here $q$ is the spatial wavenumber of the clock ripple, with wavelength
$2\pi/q$, and $b(t)$ is its dimensionless, time-dependent amplitude. The slope of a clock slice is $-bf'$ to first order and the fractional change in clock rate is $\dot b f$. Both are small. In $U_4$ we use the slice Laplacian and Gaussian, average the lapse over the four interaction points and hold $\zeta_\Sigma W$ fixed as the slice volume $W$ changes.

We write the dynamics in terms of a variable $\tau$ labeling time for the wavy slice and the variable $x$ along the slice.The tilt couples to scalar momentum through $bJ$, and the change in clock rate couples to scalar energy through $-\dot b E$. Here $J=\int f'\mathcal P_x\,\dd^3x$ and $E=\int f\mathcal H\,\dd^3x$, including the quartic in $\mathcal H$.

For clock frequencies well below any excitation gaps, and with no quartic term, the free clock is described by a  Lagrangian
\begin{equation}
 L^{(2)}=\frac V2\left[\frac{M^4}{q^2}\: \dot b^2
                   -M^4 \: b^2\right]
 \label{eq:clockrippleL}
\end{equation}
At fixed $q$ this quadratic Lagrangian is the simple harmonic approximation
for the ripple amplitude $b(t)$. 

With the quartic, this Lagrangian is modified, when we keep quadratic terms in the fluctuation, to
\begin{equation}
L^{(2)}=\frac V2\left[\left(\frac{M^4}{q^2}+D(q)\right)\dot b^2
                   -\left(M^4-K_0(q)\right)b^2\right]
 \label{eq:clockrippleLQ}
\end{equation}

We compute $K_0(q)$ and $D(q)$ by expanding the scalar vacuum response
through second order in the clock fluctuation $\chi$, keeping $U_4$ in
the scalar Hamiltonian. A static ripple shifts the scalar vacuum energy by $-VK_0b^2/2$. A time-dependent ripple also adds $VD\dot b^2/2$ to the kinetic term. For reference volume $V$ and scalar excitation gaps $\Delta_N=E_N-E_0$, the coefficients are
\begin{equation}
 \begin{aligned}
 K_0(q)&=\left[\frac2V\sum_{N\ne0}
       \frac{|\langle N|J|0\rangle|^2}{\Delta_N}\right]_{\text{interacting}-\text{free}}\\
 D(q)&=\frac2V\sum_{N\ne0}
       \frac{|\langle N|E-iJ/\Delta_N|0\rangle|^2}{\Delta_N}
 \end{aligned}
 \label{eq:clockcoefficients}
\end{equation}

The functions $K_0(q)=D(q)=0$ for the free Klein Gordon field. This is because local energy conservation gives
$\partial_t\mathcal H=\sum_i\partial_i\mathcal P_i$, where $\mathcal P_i$
is the momentum density operator.
Since $f$ depends only on $x$, integration by parts in the periodic box gives
\[
 \dot E=\int f\,\partial_t\mathcal H\,\dd^3x
       =-\int f'\mathcal P_x\,\dd^3x=-J
\]
In the free theory $\dot E=i[H_{KG},E]$, so
$\langle N|J|0\rangle=-i\Delta_N\langle N|E|0\rangle$.
The two terms in $\langle N|E-iJ/\Delta_N|0\rangle$ therefore cancel
for each $N\ne0$, giving zero free-field contribution to $D(q)$.

For $b(t)=b_0\cos(\omega t)$ its equation of motion gives
\begin{equation}
 \frac{\omega^2}{q^2}=\frac{1-K_0(q)/M^4}{1+q^2D(q)/M^4}
 \label{eq:clockdispersion}
\end{equation}
If $K_0(q)=D(q)=0$, the mode obeys
$\ddot b+q^2b=0$. The ripple is then a standing wave of the free,
massless clock field, with $\omega=q$ and phase velocity 1 (speed of light).
A positive $K_0$ weakens the restoring force, and a positive $D$ adds inertia. For $0<K_0(q)<M^4$ and finite $D(q)\ge0$, the ripple is stable and its phase speed $\omega/q$ is below 1 (subliminal).

\end{document}